\documentclass{llncs}
\usepackage{makeidx}  
\usepackage{graphicx,amssymb,amsmath}
\usepackage{stmaryrd}
\usepackage{hyperref}
\begin{document}
\frontmatter          
\pagestyle{headings}  
\addtocmark{A Language-theoretic View on Guidelines and Consistency Rules of UML} 
\mainmatter              
\title{A Language-theoretic View on Guidelines and Consistency Rules of UML}
\titlerunning{A Language-theoretic View}  
%
\author{Zhe Chen\inst{1} \and Gilles Motet\inst{1,2}}
\authorrunning{Zhe Chen and Gilles Motet}   
%
\tocauthor{Zhe Chen, Gilles Motet}
\institute{Laboratory LATTIS, INSA, University of Toulouse, \\135 Avenue de Rangueil, 31077 Toulouse, France\\
\email{zchen@insa-toulouse.fr} \and Foundation for an Industrial Safety Culture, \\
6 All\'ee Emile Monso, 31029 Toulouse, France\\
\email{gilles.motet@insa-toulouse.fr}}

\maketitle              

\newcommand*{\metacomp}{\overrightarrow{\cdot}}

\begin{abstract}
Guidelines and consistency rules of UML are used to control the degrees of freedom provided by the language to prevent faults. Guidelines are used in specific domains (e.g., avionics) to recommend the proper use of technologies. Consistency rules are used to deal with inconsistencies in models. However, guidelines and consistency rules use informal restrictions on the uses of languages, which makes checking difficult. In this paper, we consider these problems from a language-theoretic view. We propose the formalism of C-Systems, short for ``formal language control systems''. A C-System consists of a controlled grammar and a controlling grammar. Guidelines and consistency rules are formalized as controlling grammars that control the uses of UML, i.e. the derivations using the grammar of UML. This approach can be implemented as a parser, which can automatically verify the rules on a UML user model in XMI format. A
comparison to related work shows our contribution: a generic top-down and syntax-based approach that checks language level constraints at compile-time.
\end{abstract}

\section{Introduction}
\label{Sec:intro}

The UML (Unified Modeling Language) is a graphic modeling language developed by OMG (Object Management Group), and defined by the specifications \cite{UMLi} and \cite{UMLs}. UML has emerged as the software industry's dominant modeling language for specifying, designing and documenting the artifacts of systems \cite{MRR02}\cite{BRJ05}.

Evolving descriptions of software artifacts are frequently
inconsistent, and tolerating this inconsistency is important
\cite{Bal91}\cite{EC01}. Different developers construct and update
these descriptions at different times during development
\cite{NER00}, thus resulting in inconsistencies. They develop multiple
views on a system providing pieces of information which are
redundant or complementary. Constraints exist on these pieces of
information whose violation leads to inconsistent models.
Inconsistency problems of UML models have attracted great attention
from both academic and industrial communities
\cite{UMLcon02}\cite{UMLcon03}\cite{UMLcon04}. A list of 635
consistency rules are identified by \cite{VMM05a}\cite{VMM05b}.

Guidelines, which also contain a set of rules, are often required on models which are specific to a given context. For instance, OOTiA (Object-Oriented Technology in Aviation) demands that ``the length of an inheritance should be less
than 6'' \cite{FAA}. This context is domain specific. If these constraints are not respected, the presence of faults is not sure but its risk is high. The context can also be technology specific. For instance, ``multiple inheritance should be avoided in safety critical, certified systems'' (IL \#38 of \cite{FAA}), if the UML models are implemented by Java code, as this language does not provide the multiple inheritance mechanism.

It seems that consistency rules and guidelines are irrelevant at first glance. However, in fact, they have the same origin from a language-theoretical view. We noticed that both of the two types of potential faults in models come from {\em the degrees of freedom} offered by languages. These degrees of freedom cannot be eliminated without reducing the language capabilities \cite{Mot09}. For instance, the multiple diagrams in UML are useful, as they describe various viewpoints on one system, even if they are at the origin of numerous inconsistencies. In the same way, multiple inheritance can be implemented in the C++ language.

To prevent these risks of faults, {\em the use of languages} must be
controlled. To do it, guidelines are old and popular means in
industry. However, their expression is informal and their checking
is difficult. For instance, 6 months were needed to check 350
consistency rules on an avionics UML model including 116 class
diagrams.

This paper aims at formalizing {\em the acceptable use of languages} and proposing a way to check {\em the use correctness}, by considering guidelines and consistency rules from a language-theoretical view. To achieve this goal, acceptable uses of a language are defined as a grammar handling the productions of the grammar of the language. To support this idea, UML must be specified by a formal language, or at least a language with precisely defined syntax, e.g., XMI in this paper. Thus, a graphic model can be serialized. This formalism also provides a deeper view on the origin of inconsistencies in models.

This paper is organized as follows. First, we introduce the grammar of UML in XMI in Section \ref{Sec:G_UML}. Then in Section \ref{Sec:lcs}, we define the {\em C-System}, i.e. a formalism containing controlling grammars that restrict the use of the grammar of UML. We illustrate the formalism using examples in Section \ref{Sec:exmp}. Related work and implementation of this approach are discussed in Sections \ref{Sec:rw} and \ref{Sec:disc}. Section \ref{Sec:conclusion} concludes the paper.

\section{The Grammar of UML in XMI}
\label{Sec:G_UML}

XMI (XML Metadata Interchange) \cite{UML-XMI} is used to facilitate
interchanging UML models between different modeling tools in XML
format. Many tools implement the conversion, e.g., Altova
UModel$^\circledR$ can export UML models as XMI files.

A UML model in XMI is an XMI-compliant XML document that conforms to
its XML schema, and is a derivative of the {\em XMI document productions}
which is defined as a grammar. The XML schema is a derivative of the
{\em XMI schema productions}. The XMI specification defines both the XMI
schema productions and the XMI document productions in
\cite{UML-XMI}.

XMI provides a mapping between a UML user model and an XML document,
and a mapping between UML (also MOF) and an XML Schema. XMI
generates an XML file using the XMI document productions, and
generates an XML schema using the XMI schema productions. Each of
the two sets of productions composes a context-free grammar in EBNF
\cite{UML-EBNF}. A UML user model can be expressed using an
XMI-compliant XML document that conforms to the corresponding XML
Schema, and is a derivative of the XMI document grammar.

The grammar and its productions for deriving XMI-compliant XML
documents of UML models are defined in \cite{UML-XMI}. The main part
of the grammar is given here after. To make our presentation more
concise, we omit declaration and version information of XML files
(and the related productions whose names start with ``1'').

To make later reasoning easier, we modified some representations of the productions, but without changing the generative power of the grammar.

1. The choice operator ``$|$'' is used to compose several productions with the same left-hand side into a single line in \cite{UML-XMI}. We decomposed some of these productions into several productions without the choice operator. An original production $n$ having $k$ choices might be divided into a set of productions $\{ n\_ i\}_{1\leq i \leq k}$. For example, the original production 2 with three choices was divided into the productions 2\_1, 2\_2 and 2\_3.

2. The closure operator ``*'' is used to simplify the representation of the grammar in \cite{UML-XMI}, but it also would make the representation of reasoning confusing. Thus, the productions whose names start with ``3'' were added to replace the productions with closure operators.

The \textbf{grammar $G$} of UML in XMI includes the following productions
(each production is labeled by a name starting with a digit):

  \begin{verbatim}
 3_1:  XMIElements ::= 2:XMIElement
 3_2:  XMIElements ::= 2:XMIElement 3:XMIElements

 2_1:  XMIElement ::= 2a:XMIObjectElement
 2_2:  XMIElement ::= 2b:XMIValueElement
 2_3:  XMIElement ::= 2c:XMIReferenceElement

 2a_1: XMIObjectElement ::= "<" 2k:QName 2d:XMIAttributes "/>"
 2a_2: XMIObjectElement ::= "<" 2k:QName 2d:XMIAttributes ">"
                             3:XMIElements "</" 2k:QName ">"

 2b_1: XMIValueElement ::= "<" xmiName ">" value "</" xmiName ">"
 2b_2: XMIValueElement ::= "<" xmiName "nil=`true'/>"

 2c_1: XMIReferenceElement::= "<" xmiName 2l:LinkAttribs "/>"
 2c_2: XMIReferenceElement::= "<" xmiName 2g:TypeAttrib
                              2l:LinkAttribs "/>"

 2d_1: XMIAttributes ::= 2g:TypeAttrib  2e:IdentityAttribs
                         3h:FeatureAttribs
 2d_2: XMIAttributes ::= 2e:IdentityAttribs  3h:FeatureAttribs

 2e:   IdentityAttribs ::= 2f:IdAttribName "=`" id "'"

 2f_1: IdAttribName ::= "xmi:id"
 2f_2: IdAttribName ::= xmiIdAttribName

 2g:   TypeAttrib ::= "xmi:type=`" 2k:QName "'"

 3h_1: FeatureAttribs ::= 2h:FeatureAttrib
 3h_2: FeatureAttribs ::= 2h:FeatureAttrib  3h:FeatureAttribs

 2h_1: FeatureAttrib ::= 2i:XMIValueAttribute
 2h_2: FeatureAttrib ::= 2j:XMIReferenceAttribute

 2i:   XMIValueAttribute ::= xmiName "=`" value "'"

 2j:   XMIReferenceAttribute ::= xmiName "=`" (refId | 2n:URIref)+"'"

 2k:   QName ::= "uml:" xmiName  |  xmiName

 2l:   LinkAttribs ::= "xmi:idref=`" refId "'" | 2m:Link

 2m:   Link ::= "href=`" 2n:URIref "'"

 2n:   URIref ::= (2k:QName)? uriReference
  \end{verbatim}

In the grammar, the symbol ``::='' stands for the conventional
rewriting symbol ``$\rightarrow$'' in formal languages theory \cite{HU79}.
Each nonterminal starts with a capital letter, prefixing a label of
the related production, e.g., ``2:XMIElement'' is a nonterminal with
possible productions ``2\_1, 2\_2, 2\_3''. Each terminal starts with
a lowercase letter or is quoted.

\begin{figure}[tbh]
  \centering
\begin{minipage}[t]{0.4\textwidth}
  \centering
  \includegraphics[bb= 0 0 150 150, scale=1]{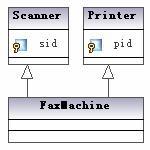}
  \caption{A Class Diagram}
  \label{Fig:ClassDiag}
\end{minipage}
\begin{minipage}[t]{0.5\textwidth}
  \centering
  \includegraphics[bb= 0 0 200 90, scale=1]{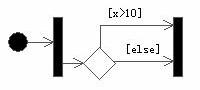}
  \caption{An Activity Diagram}
  \label{Fig:ActDiag}
\end{minipage}
\end{figure}

As an example to illustrate the use of the grammar, Figure
\ref{Fig:ClassDiag} represents a package {\em Root} which includes
three classes, where the class {\em FaxMachine} is derived from {\em
Scanner} and {\em Printer}. The core part of the exported XMI 2.1
compliant file (using Altova UModel$^\circledR$) is as follows:

\begin{verbatim}
 <uml:Package xmi:id="U00000001-7510-11d9-86f2-000476a22f44"
              name="Root">
      <packagedElement xmi:type="uml:Class"
              xmi:id="U572b4953-ad35-496f-af6f-f2f048c163b1"
              name="Scanner" visibility="public">
           <ownedAttribute xmi:type="uml:Property"
              xmi:id="U46ec6e01-5510-43a2-80e9-89d9b780a60b"
           name="sid" visibility="protected"/>
      </packagedElement>
      <packagedElement xmi:type="uml:Class"
              xmi:id="Ua9bd8252-0742-4b3e-9b4b-07a95f7d242e"
              name="Printer" visibility="public">
           <ownedAttribute xmi:type="uml:Property"
              xmi:id="U2ce0e4c8-88ee-445b-8169-f4c483ab9160"
           name="pid" visibility="protected"/>
      </packagedElement>
      <packagedElement xmi:type="uml:Class"
              xmi:id="U6dea1ea0-81d2-4b9c-aab7-a830765169f0"
              name="FaxMachine" visibility="public">
           <generalization xmi:type="uml:Generalization"
              xmi:id="U3b334927-5573-40cd-a82b-1ee065ada72c"
              general="U572b4953-ad35-496f-af6f-f2f048c163b1"/>
           <generalization xmi:type="uml:Generalization"
              xmi:id="U86a6818b-f7e7-42d9-a21b-c0e639a4f716"
              general="Ua9bd8252-0742-4b3e-9b4b-07a95f7d242e"/>
      </packagedElement>
 </uml:Package>
\end{verbatim}

This text is a derivative of the XMI document productions, c.f. the
previous grammar $G$. We may use the sequence of productions
``2a\_2, 2k(Package), 2d\_2, 2e, 2f\_1, 3h\_1, 2h\_1, 2i'' to derive
the following sentential form:

\begin{verbatim}
 <uml:Package xmi:id="U00000001-7510-11d9-86f2-000476a22f44"
              name="Root">
      3:XMIElements "</" 2k:QName ">"
\end{verbatim}

Note that the production 2k has a parameter {\em xmiName}, i.e. the
value of the terminal when apply the production. In a derivation, we
specify a value of the parameter as ``2k(value)''. For example,
``2k(Package)'' is a derivation using 2k with {\em xmiName} =
``Package''. For simplicity, we consider ``2k(value)'' as a terminal
as a whole. We continue to apply productions, and finally derive the
XMI file previously presented.

Notice that the model of Fig. \ref{Fig:ClassDiag} (both in UML and XML) does not conform to the guidelines in OOTiA about multiple inheritance, since it uses multi-inheritance. The model of Fig. \ref{Fig:ActDiag} has an
inconsistency: ``the number of outgoing edges of $ForkNode$ is not
the same as the number of incoming edges of $JoinNode$''. In
particular, $JoinNode$ joins two $outgoing$ edges from the same
$DecicionNode$, This join transition will never be activated, since
only one of the two $outgoing$ edges will be fired.

We will define a formal model to check the conformance to these
rules by controlling the use of the grammar of UML.

\section{The C-System: A Formal Language Control System}
\label{Sec:lcs}

In this section, we propose the formal model for controlling the use
of grammars based on classical language theory \cite{HU79}.

Let $G = (N, T, P, S)$ be a grammar, where $N$ is the set of
nonterminals, $T$ is the set of terminals, $P$ is the set of
productions of the form $l:A \rightarrow \alpha$ where $l$ is the
name of the production, $A \in N$, $\alpha \in (N \bigcup T)^*$, and $S$ is
the start symbol. A derivation using a specified production $p$ is
denoted by $\alpha \overset{p}\Rightarrow \beta$, and multiple
derivations are denoted by $\alpha \Rightarrow^* \gamma$.

\begin{definition}
A \textbf{controlling grammar} $\hat{G}$ over a \textbf{controlled
grammar} (or simply grammar) $G = (N, T, P, S)$ is a quadruple
$\hat{G}=(\hat{N}, \hat{T}, \hat{P}, \hat{S})$, where $\hat{T}=P$.
The language $L(\hat{G})$ is called a \textbf{controlling language}.
\hfill$\Box$
\end{definition}

The symbol $\hat{G}$ is read ``control $G$'' or ``$G$ hat''. For
making reading easier, we assume that $N\cap\hat{N}=\emptyset$.
$\hat{T}=P$ means that the terminals of $\hat{G}$ are exactly the
productions of $G$.

If we use an automaton $A$ to process the input string, such that
$L(A)=L(G)$, then we can also use a \textbf{controlling automaton}
$\hat{A}$ to represent the controlling language.

As we know, each string $w \in L(G)$ has at least one {\em leftmost} derivation (denoted by ``$lm$'') using a sequence of productions from $P$, e.g. $p_1p_2 ... p_k$. The controlling grammar restricts the derivation in the sense that the sequences of applied productions should be in the language it specifies, i.e., $p_1p_2 ... p_k \in L(\hat{G})$. Formally, we have the following definition.

\begin{definition}\label{Def:LCS}
Given a grammar $G=(N, T, P, S)$, the language of the grammar with a
controlling grammar $\hat{G}$ is:
\begin{center}
$L(G\metacomp\hat{G}) = \{w | S
\underset{lm}{\overset{p_1}\Rightarrow} w_1
\cdots \underset{lm}{\overset{p_k}\Rightarrow} w_k = w$, $p_1,p_2,...,p_k \in P$ and
$p_1p_2...p_k \in L(\hat{G}) \}$
\end{center}
We say that $G$ and $\hat{G}$ constitute a \textbf{C-System} $C = G\metacomp \hat{G}$, short for \textbf{formal language control system}. The language $L(C) = L(G\metacomp\hat{G})$ is called a \textbf{global system language}.\hfill$\Box$
\end{definition}

The symbol $\metacomp$ is called ``meta composition''. Its left operand is controlled by the right operand, which is a meta level grammar. If we use automata-based notations, a string $w \in L(A\metacomp\hat{A})$ if and only if $A$ accepts $w$, and $\hat{A}$ accepts the sequence of the labels of the transitions used.

A \textbf{\emph{regular C-System}} is a C-System of which the controlled grammar $G$ is a regular grammar (or $A$ is a finite automaton). Some variants of regular C-Systems are proposed for ensuring system safety requirements, e.g. Input/Output C-Systems \cite{CM09a}, Interface C-Systems \cite{CM09d}\cite{CM09b}. We denote by $\mathcal{C}_R$ the family of regular C-Systems.

A \textbf{\emph{context-free C-System}} is a C-System of which the controlled grammar $G$ is a context-free grammar (or $A$ is a pushdown automaton). We denote by $\mathcal{C}_{CF}$ the family of context-free C-Systems.

Generally, we denote by $\mathcal{C}_X^Y$ the family of C-Systems that consist of $X$-type controlled grammar and $Y$-type controlling grammar, where $X,Y \in \{R, CF\}$. Although $X,Y$ could be also other types in Chomsky hierarchy, e.g. context-sensitive, this is beyond the scope of this paper.

Obviously, the set of accepted inputs is a subset of the controlled
language, such that the sequence of the applied productions belongs
to the controlling language. Consider a simple example as follows.

\begin{example}\label{Exmp:Gr_Gr}
Given a regular grammar $G$ and a regular controlling grammar $\hat{G}$:

\begin{equation*}
G \left\{ \begin{aligned}
                p_1:\ & S \rightarrow aS\\
                p_2:\ & S \rightarrow bS\\
                p_3:\ & S \rightarrow \epsilon
              \end{aligned} \right.~~~~
\hat{G} \left\{ \begin{aligned}
                \hat{S} &\rightarrow p_1 \hat{S} | p_3 \hat{S} | p_2 A  \\
                A &\rightarrow p_2A | p_3A | \epsilon  \\
              \end{aligned} \right.
\end{equation*}

$L(G)$ accepts the language $(a|b)^*$, e.g., $aab$, $abab$.
$L(\hat{G})$ accepts the language $(p_1|p_3)^* p_2 (p_2|p_3)^*$. The
trivial grammar $G$ is considered to provide a simple illustration
of the introduced principles.

The grammars $G$ and $\hat{G}$ constitute a regular C-System $C = G \metacomp \hat{G} \in \mathcal{C}_R^R$.

Given the string $aab \in L(G)$, we conclude that $aab \in
L(G\metacomp\hat{G})$, because we have the leftmost derivations $S
\overset{p_1}\Rightarrow aS \overset{p_1}\Rightarrow aaS
\overset{p_2}\Rightarrow aabS \overset{p_3}\Rightarrow aab$, where
$p_1p_1p_2p_3\in L(\hat{G})$ as $\hat{S} \Rightarrow p_1\hat{S}
\Rightarrow p_1p_1\hat{S} \Rightarrow p_1p_1p_2A \Rightarrow
p_1p_1p_2p_3A \Rightarrow p_1p_1p_2p_3$. On the contrary, we have
$abab \not\in L(G\metacomp\hat{G})$. Although we have the leftmost
derivation $S \overset{p_1}\Rightarrow aS \overset{p_2}\Rightarrow
abS \overset{p_1}\Rightarrow abaS \overset{p_2}\Rightarrow ababS
\overset{p_3}\Rightarrow abab$, $p_1p_2p_1p_2p_3 \not\in
L(\hat{G})$.

In fact, the language $L(C) = L(G\metacomp\hat{G})$ is equivalent to the
language $a^*b^+$, which is the subset of $(a|b)^*$ satisfying the
constraints: ``every $a$ should appear before $b$'' and ``at least
one $b$''. \hfill$\Box$
\end{example}

We remark here that our model is different from regularly controlled grammars \cite{GS68}\cite{DPS97}, in the sense that we restrict derivations to be {\em leftmost} and allow {\em context-free controlling grammars}. These differences result in different theoretical results, which are beyond the scope of this paper.

\section{Examples}
\label{Sec:exmp}

In this section we use some practical examples to illustrate the
idea of the previous section. We denote the grammar of UML by $G =
(N,T,P,S)$, where $P$ is the set of productions listed in Section
\ref{Sec:G_UML}, and each production $p \in P$ is labeled by a name
starting with a digit.

\begin{example}
Consider two rules on class diagrams:

\textbf{Rule 1:} Each class can have at most one generalization.
This rule is a guideline, as we mentioned in Section 1 and at the end of Section 2. This rule is
also a consistency rule in the context of Java, since Java does not
allow multiple inheritance. However we may derive a class from
multiple classes in the context of C++.

\textbf{Rule 2:} Each class can have at most 30 attributes. This
rule may be adopted by software authorities as a guideline in
avionics, in order to increase the safety of software systems by
minimizing the complexity of classes.
\end{example}

Note that these rules cannot be explicitly integrated into the
grammar of UML, but only recommended as guidelines or consistency
rules. We cannot put rule 1 into the standard of UML, since UML
models can be implemented with both C++ and Java programming languages.
Rule 2 is a restriction for a specific domain, and we should not
require all programmers to use limited attributes by specifying the
UML language.

We aim to specify the rules from the meta-language level, thus
control the use of the language. Consider the example of Fig.
\ref{Fig:ClassDiag}, to obtain the associated XMI text, the sequence
of applied productions of $G$ in the leftmost derivation is as
follow (``...'' stands for some omitted productions, to save space):

\begin{verbatim}
 2a_2, 2k(Package), 2d_2, 2e, 2f_1, 3h_1, 2h_1, 2i,
 ..., 2k(packagedElement), ..., 2k(Class),
       ..., 2k(ownedAttribute), ..., 2k(Property),
 ..., 2k(packagedElement),
 ..., 2k(packagedElement), ..., 2k(Class),
       ..., 2k(ownedAttribute), ..., 2k(Property),
 ..., 2k(packagedElement),
 ..., 2k(packagedElement), ..., 2k(Class),
       ..., 2k(generalization), ..., 2k(Generalization),
       ..., 2k(generalization), ..., 2k(Generalization),
 ..., 2k(packagedElement),
 ..., 2k(Package)
\end{verbatim}

Let $c,g$ stand for $2k(Class), 2k(Generalization)$, respectively.
Note that the occurrence of two $g$ after the third $c$ violates
Rule 1. In fact, all the sequences of productions in the pattern
``$...c...g...g...$'' are not allowed by the rule (there is no $c$
between the two $g$), indicating that the class has two
generalizations.

Thus, we propose the following controlling grammar $\hat{G}_c$ to
restrict the use of the language to satisfy Rule 1:
\begin{equation} \label{eq:G_c}
\hat{G}_c \left\{ \begin{aligned}
S   &\rightarrow c~~Q_c ~|~ D~S ~|~ D \\
Q_c &\rightarrow c~~Q_c ~|~ g~~Q_g ~|~ D~Q_c ~|~ D\\
Q_g &\rightarrow c~~Q_c ~|~ D~Q_g ~|~ D\\
D   &\rightarrow \{ p~|~p \in P ~\wedge ~ p \not\in \{c,g\}\}
\end{aligned} \right.
\end{equation}
where $S, Q_c, Q_g, D$ are nonterminals, $D$ includes all
productions except $c, g$. $L(\hat{G}_c)$ accepts the sequences of
productions satisfying Rule 1.

\begin{figure}
  \centering
  \includegraphics[scale=1]{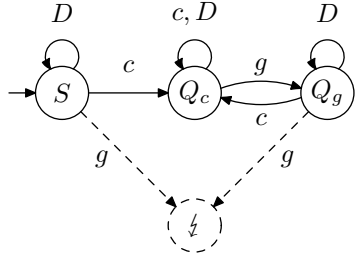}\\
  \caption{The Automaton $\hat{A}_c$}\label{Fig:A_G_c}
\end{figure}

Implicitly, the controlling grammar specifies an automaton
$\hat{A}_c$ in Fig. \ref{Fig:A_G_c}, where $\lightning$ is an
implicit error state (the dashed circle). Strings of the pattern
$D^*cD^*gD^*gD^*$ will lead $\hat{A}_c$ to the error state.

If the sequence of productions applied to derive a model is accepted
by the language $L(\hat{G}_c)$, then the model conforms to Rule 1.
In Fig. \ref{Fig:ClassDiag}, the derivation of the class
$FaxMachine$ uses the pattern $D^*cD^*gD^*gD^* \not\in
L(\hat{G}_c)$, which leads to $\lightning$ of the automaton, thus it
violates Rule 1. On the contrary, the derivations of $Scanner$ and
$Printer$ are accepted by $L(\hat{G}_c)$, thus satisfy Rule 1.

Now consider Rule 2. Let $c,pr,pe$ stand for $2k(Class)$,
$2k(Property)$, \\$2k(PackagedElement)$, respectively. Note that the
occurrence of more than 30 $pr$ after a $c$ violates Rule 2. In
fact, all the sequences of productions in the pattern
``$...c...(pr...)^n, n > 30$'' are not allowed by the rule (there is
no $c$ between any two $pr$), indicating that the class has more
than 30 attributes.

To satisfy Rule 2, we propose the following controlling grammar
$\hat{G}_p$ to restrict the use of the language:

\begin{equation} \label{eq:G_p}
\hat{G}_p \left\{ \begin{aligned}
S   &\rightarrow pe~~S ~|~ c~~Q_c ~|~ D~S ~|~ D\\
Q_c &\rightarrow pe~S ~|~ c~~Q_c ~|~ pr~~Q_1 ~|~ D~Q_c ~|~ D\\
Q_i &\rightarrow pe~S ~|~ c~~Q_c ~|~ pr~~Q_{i+1} ~|~ D~Q_i ~|~ D ~~(1\leq i < 30) \\
Q_{30} &\rightarrow pe~~S ~|~ c~~Q_c ~|~ D~Q_{30} ~|~ D \\
D   &\rightarrow \{ p~|~p \in P ~\wedge ~ p \not\in \{c,pr,pe\}\}
\end{aligned} \right.
\end{equation}
where $S, Q_c, Q_i$ are nonterminals, $D$ includes all productions
except $c, pr, pe$. $L(\hat{G}_p)$ accepts the sequences of
productions satisfying Rule 2.

\begin{figure}
  \centering
  \includegraphics[scale=1]{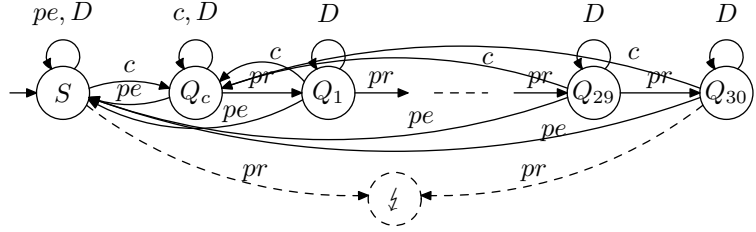}\\
  \caption{The Automaton $\hat{A}_p$}\label{Fig:A_G_p}
\end{figure}

Implicitly, the controlling grammar specifies an automaton
$\hat{A}_p$ in Fig. \ref{Fig:A_G_p}. Strings of the pattern
``$D^*cD^*(pr~D^*)^n, n>30$'' will lead $\hat{A}_p$ to the error
state.

If the sequence of productions applied to derive a model is accepted
by the language $L(\hat{G}_p)$, then the model conforms to Rule 2.
In Fig. \ref{Fig:ClassDiag}, the derivations of the classes
$Scanner$ and $Printer$ use the pattern $D^*cD^*prD^* \in
L(\hat{G}_p)$, thus satisfy Rule 2.

Thanks to the controlling grammars, when a model violates required
rules, the controlling language will reject the model (an implicit
error state $\lightning$ will be activated). Some error handling
method may be called to process the error, e.g., printing an error
message indicating the position and the cause.

We can also use controlling grammar to handle a consistency rule
concerning activity diagrams.

\begin{example}\label{Exmp:Act_Dig}
In an activity diagram, the number of outgoing edges of $ForkNode$
should be the same as the number of incoming edges of its pairwise
$JoinNode$.
\end{example}

Let $n,f,j,i,o$ stand for $2k(node)$, $2k(ForkNode)$,
$2k(JoinNode)$, $2k(incoming)$, $2k(outgoing)$, respectively. We
propose the following controlling grammar $\hat{G}_a$ to restrict
the use of the language to satisfy the rule:

\begin{equation} \label{eq:G_a}
\hat{G}_a \left\{ \begin{aligned}
S   &\rightarrow N~~F~~I^*~~Q~~O^*~~N ~|~ N~~I^*~~O^*~~N ~|~ D^*\\
Q   &\rightarrow O~~Q~~I ~|~ N~~S~~N~~J \\
N   &\rightarrow n~~D^*\\
F   &\rightarrow f~~D^*\\
J   &\rightarrow j~~D^* \\
I   &\rightarrow i~~D^* \\
O   &\rightarrow o~~D^* \\
D   &\rightarrow \{ p~|~p \in P ~\wedge ~ p \not\in \{n,f,j,i,o\}\}
\end{aligned} \right.
\end{equation}

$L(\hat{G}_a)$ accepts all the sequences of productions of the
pattern \\$NFI^*O^nN S NJI^nO^*N$, which leads to models respecting
the rule. This context-free grammar implicitly specifies a PDA
(Pushdown Automaton \cite{HU79}), which is more complex than the
automata in Figures \ref{Fig:A_G_c} and \ref{Fig:A_G_p}.

Globally, any UML user model $M$ derived from the C-System $C = G \metacomp \hat{G}_a \in \mathcal{C}_{CF}^{CF}$, i.e. $M \in L(C)$, conforms to the rule in Example \ref{Exmp:Act_Dig}.

As a more concrete instance, we consider the model in Fig.
\ref{Fig:ActDiag}. The XMI-compliant document of the model in Fig.
\ref{Fig:ActDiag} is the follows:
\begin{verbatim}
 <packagedElement xmi:type="uml:Activity"
              xmi:id="U937506ed-af64-44c6-9b4c-e735bb6d8cc6"
              name="Activity1" visibility="public">
 <node xmi:type="uml:InitialNode" xmi:id="U16aa15e8-0e5d-
        4fd1-930a-725073ece9f0">
    <outgoing xmi:idref="Ue9366b93-a45b-43f1-a201-2038b0bd0b30"/>
 </node>
 <node xmi:type="uml:ForkNode" xmi:id="U26768518-a40c-
                   4713-b35e-c267cc660508" name="ForkNode">
    <incoming xmi:idref="Ue9366b93-a45b-43f1-a201-2038b0bd0b30"/>
    <outgoing xmi:idref="Ua800ba9b-e167-4a7c-a9a9-80e6a77edeb7"/>
 </node>
 <node xmi:type="uml:DecisionNode" xmi:id="Uc9e4f0de-8da6-
        4c98-9b95-b4cde30ccfc0" name="DecisionNode">
    <incoming xmi:idref="Ua800ba9b-e167-4a7c-a9a9-80e6a77edeb7"/>
    <outgoing xmi:idref="Ua4a2b313-13d6-4d69-9617-4803560731ef"/>
    <outgoing xmi:idref="U6eede33f-98ac-4654-bb17-dbe6aa7e46be"/>
 </node>
 <node xmi:type="uml:JoinNode" xmi:id="Ud304ce3c-ebe4-
                   4b06-b75a-fa2321f8a151" name="JoinNode">
    <incoming xmi:idref="Ua4a2b313-13d6-4d69-9617-4803560731ef"/>
    <incoming xmi:idref="U6eede33f-98ac-4654-bb17-dbe6aa7e46be"/>
 </node>
 <edge xmi:type="uml:ControlFlow"
        xmi:id="Ua4a2b313-13d6-4d69-9617-4803560731ef"
        source="Uc9e4f0de-8da6-4c98-9b95-b4cde30ccfc0"
        target="Ud304ce3c-ebe4-4b06-b75a-fa2321f8a151">
    <guard xmi:type="uml:LiteralString"
        xmi:id="U6872f3b3-680c-430e-bdb3-21c0a317d290"
        visibility="public" value="x>10"/>
 </edge>
 <edge xmi:type="uml:ControlFlow"
        xmi:id="U6eede33f-98ac-4654-bb17-dbe6aa7e46be"
        source="Uc9e4f0de-8da6-4c98-9b95-b4cde30ccfc0"
        target="Ud304ce3c-ebe4-4b06-b75a-fa2321f8a151">
    <guard xmi:type="uml:LiteralString"
        xmi:id="Ub853080d-481c-46ff-9f7c-92a31ac24349"
        visibility="public" value="else"/>
 </edge>
 <edge xmi:type="uml:ControlFlow"
        xmi:id="Ua800ba9b-e167-4a7c-a9a9-80e6a77edeb7"
        source="U26768518-a40c-4713-b35e-c267cc660508"
        target="Uc9e4f0de-8da6-4c98-9b95-b4cde30ccfc0"/>
 <edge
        xmi:type="uml:ControlFlow"
        xmi:id="Ue9366b93-a45b-43f1-a201-2038b0bd0b30"
        source="U16aa15e8-0e5d-4fd1-930a-725073ece9f0"
        target="U26768518-a40c-4713-b35e-c267cc660508"/>
 </packagedElement>
\end{verbatim}

It is easy to detect that the sequence of applied productions\\
``$...nD^*fD^*iD^*oD^*nD^*$ ... $nD^*jD^*iD^*i...$'' is not accepted
by $L(\hat{G}_a)$ (one $o$ follows $f$, while two $i$ follow $j$),
thus there is an inconsistency.

We remark here that there are two preconditions of using the
controlling grammar concerning the sequences of the model elements
in the XML document: 1. $ForkNode$ must appear before its pairwise
$JoinNode$; 2. $incoming$ edges must appear before $outcoming$ edges
in a node. The two conditions are trivial, since it is easy to
control their positions in the exporting XMI documents in implementing
such a transformation.

\section{Related Work}
\label{Sec:rw}

The most popular technique for verifying software correctness is {\em model checking} \cite{CGP00}. In this framework, we have three steps in verifying a system. First, we formalize system behavior as a model (e.g., a transition system, a Kripke model \cite{HR04}). Second, we specify the properties that we aim at validating using temporal logics. Third, we use a certain checking algorithm to search for a counterexample which is an execution trace violating the specified properties. If the algorithm finds such a counterexample, we have to correct the original design.

Most checking tools use specific semantics of UML diagrams. They
have the flavor of model checking, e.g., Egyed's UML/Analyzer
\cite{Egy07a}\cite{Egy07b} and OCL (Object Constraint Language)
\cite{UML-OCL}. At first, developers design UML diagrams as a model.
Then, we specify the consistency rules as OCL or similar
expressions. Certain algorithms are executed to detect
counterexamples that violate the rules \cite{CPC04}. Note that these
techniques do not discriminate the rules on the model level and
those concerning the language level features.

Unlike these techniques, our framework takes another way of ensuring
correctness. It consists of the following steps:
\begin{enumerate}
  \item Specifying the grammar $G$ of a language. It specifies an {\em operational
semantics}, which defines what a language is able to model.
Developing the grammar is mainly performed by {\em language
designers}.
  \item Modeling correctness rules of the use of languages as a controlling grammar $\hat{G}$.
  It specifies a {\em correctness semantics}, which defines
what a language is authorized to derive. This process is the duty of
{\em safety engineers} whose responsibility is to assure the correct
use of the language.
  \item The two grammars constitute a consistent language as a whole, that is,
  any derivations of the global system language is a correct and consistent use of the language.
\end{enumerate}

\begin{figure}
  \centering
  \includegraphics{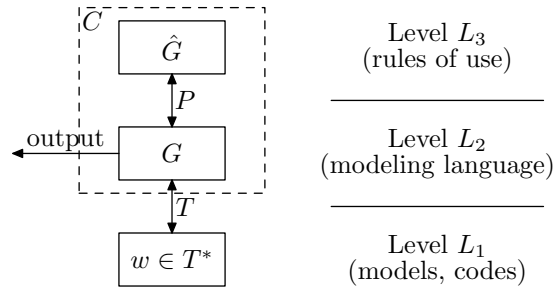}\\
  \caption{Three Levels of the Framework}\label{Fig:3level}
\end{figure}

In particular, our work differs from model checking in the following
aspects:

1. Our work has different objectives, and uses different approaches
to those of model checking. As we show in Fig. \ref{Fig:3level},
model checking techniques use a \textbf{bottom-up approach} --- they
verify execution traces $T^*$ at the lower level $L_1$ to prove the
correct use of the grammar $G$ at the middle level $L_2$. Whereas
our proposal uses a \textbf{top-down approach}
--- we model correctness rules as acceptable sequences of
productions ($P^*$) at the higher level $L_3$ to ensure the correct
use of $G$. Then any derivatives (at $L_1$) that conform to the
C-System $C = G \metacomp \hat{G}$ are definitely a correct use. So the two
techniques are complementary.

2. Our work and model checking express language-level and
model-level constraints, respectively. Language-level constraints
are more effective, because they implicitly have reusability. That
is, we only need to develop one language-level constraint and apply
it to all the models in the language. However, using model checking,
we need to replicate model-level constraints for each model.
Additionally, model checking can process model-specific constraints.

3. Our work and model checking use syntax-based and semantics-based approaches (or static and dynamic analysis), respectively. As a result, our approach is generic and metamodel-independent, and concerns little about semantics. So it can be applied to all MOF-compliant languages, not only to UML. However, model checking techniques depends on the semantics of a language, thus specific algorithms should be developed for different models.

4. Our work and model checking catch errors at compile-time and runtime, respectively. As a result, our approach implements membership checking of context-free languages, which is decidable. That is, it searches in a limited space, which is defined by grammars. Model checking may search in a larger, even infinite space, so we have to limit the
space of computing, and introduce the risk of missing solutions.

\section{Discussion}
\label{Sec:disc}

In this section, we would like to shortly discuss some issues which are beyond the scope of this paper.

The first issue concerns the implementation of controlling grammars.
The controlled and controlling grammars can be implemented using two
parsers separately. The technique for constructing a parser from a
context-free grammar is rather mature \cite{Ear70}\cite{ASU86}. Some
tools provide automated generation of parsers from a grammar
specification, such as Yacc, Bison.

Notice that the inputs of controlling parsers are the sequences of
productions applied in the parsing of $L(G)$. So there are
communications between the two parsers. Once module $G$ uses a
production $p_i$, then the name of the production is sent to
$\hat{G}$ as an input. If $L(\hat{G})$ accepts the sequence of
productions and $L(G)$ accepts the model, then $L(G \metacomp
\hat{G})$ accepts the model.

The second issue deals with multiple rules. If we have multiple guidelines or consistency rules, each rule is formalized using a grammar. We can develop an automated tool that converts the grammars into automata, and then combine these automata to compute an intersection, i.e., an automaton $A'$ \cite{HU79}. The intersection $A'$ can be used as a controlling automaton, which specifies a controlling language $L(A')$ that includes all the semantics of the rules.

The third issue is about the tradeoff between cost and benefits of applying the proposed approach. It seems that writing a controlling grammar is expensive, because it involves formal methods. However, it is probably not the case. As we mentioned, a controlling grammar specify language-level constraints, and can be reused by all the models derived from the controlled grammar. Thus the controlling grammar can be identified and formalized by the organizations who define the language or its authorized usage, e.g., OMG and FAA (Federal Aviation Administration), respectively. Developers and software companies can use the published standard controlling grammar for checking inconsistencies in their models. By contraries, if every user writes their own checking algorithms and codes, e.g., in OCL or other programming languages, the codes will be hard to be reused by other users who have different models to check. Thus the total cost of all the users may be higher. Of course, more empirical results on the tradeoff is a good direction for future work.

\section{Conclusion}
\label{Sec:conclusion}

We provided a language-theoretic view on guidelines and consistency rules of UML. We proposed the formalism of {C-Systems}, short for ``formal language control systems''. To the best of our knowledge, none related work proposed similar methodologies. Rules are considered as controlling grammars which control the use of modeling languages. This methodology is generic, syntax-based and metamodel-independent. It provides a top-down approach that checks and reports violations of language level constraints at compile-time. It can be also applied to all MOF-compliant languages, not only to UML, since it does not depend on the specific semantics of languages.

Since we focused on the methodological foundation, one of the future
work is to develop an automated checking tool implementing the
presented principles. We will also examine instant checking
techniques of our method. One feature of UML/Analyzer is instant
checking, which only verifies the small portion where the model
changes, in order to save the cost of checking \cite{Egy06}.
Intuitively, our approach is also easy to be extended to instant
checking. We only need to generate the XMI document of the changed
part of diagrams (e.g. a class in a class diagram), and verify it.
However, this calls for more works in detail.

\bibliographystyle{splncs}
\bibliography{main}

\end{document}